\documentclass[twocolumn]{jetpl}
\usepackage{amsmath,amssymb,pgf,graphicx,color}
\usepackage{hyperref}
\hypersetup{colorlinks=false}

\newcommand{\be}{\begin{equation}}
\newcommand{\ee}{\end{equation}}
\newcommand{\bea}{\begin{eqnarray}}
\newcommand{\eea}{\end{eqnarray}}

\title{Prospects for strangelet detection with large-scale cosmic ray observatories}
\rtitle{Strangelet detection with cosmic ray observatories.}
\sodtitle{Prospects for strangelet detection with large-scale cosmic ray observatories}
\author{M.\,S.\,Pshirkov$^{+*\times}$\/\thanks{e-mail: pshirkov@sai.msu.ru}}
\rauthor{M.\,S.\,Pshirkov}
\sodauthor{Pshirkov}
\address{$^+$ Sternberg Astronomical Institute, Lomonosov Moscow State University, Universitetsky prospekt 13, 119992, Moscow, Russia\\~\\
$^*$Institute for Nuclear Research of the Russian Academy of Sciences, 117312, Moscow, Russia\\~\\
$^\times$Pushchino Radio Astronomy Observatory, 142290, Pushchino, Russia
}
\abstract{
Quark matter which contains s-quarks in addition to u- and d- could be stable or metastable. In this case, lumps made of this strange matter, called strangelets, could occasionally hit the Earth. When travelling through the atmosphere they would behave not dissimilar to usual high-velocity meteors with only exception that, eventually, strangelets reach the surface.
As these encounters are expected to be extremely rare events, very large exposure is needed for their observation.
Fluorescence detectors utilized in large ultra-high energy cosmic ray observatories, such as the Pierre Auger observatory and the Telescope Array are well suited for a task of the detection of these events.
The flux limits that can be obtained with the Telescope Array fluorescence detectors could be as low as $5\times10^{-22}~\mathrm{cm^{-2} s^{-1} sr^{-1}}$ which would improve by 1.5 orders of magnitude the strongest present limits obtained from ancient mica crystals. 
}
\begin{document}

\maketitle

{\bf 1. Introduction.}

True ground state of nuclear matter could be realized in ultra-dense lumps consisting of u-, d-, s- quarks, so-called ``strange'' matter \cite{Bodmer1971,Witten1984}. Though properties of this phase of matter is not exactly known, calculations show that these lumps  could be stable  in extremely wide range of baryon numbers $A$: $O(1)<A<10^{57}$. When these clumps are of stellar masses they are usually called 'strange stars', whereas somewhat lighter clumps are dubbed  strangelets.  
Strangelets could have been produced in the very early Universe, though only the most massive ones could survive to the present time. The compact objects akin to neutron stars could be 'strangelet factories' now: some collapsing objects could transgress the stage of usual neutron stars and turn directly into the strange stars. Moreover, even tiny amount of strange matter could convert usual neutron star into a strange one (see below). Some compact binaries, coalescing less than in the Hubble time, could eject up to 0.1 $M_{\odot}$ at the final stage of merging, if some of these stars had been strange, the ISM would be thoroughly polluted with strangelets \cite{Madsen1999,Madsen2006}.  Also, the strangelets could possibly be created in the heavy-ion collision experiments, and because of that they sometimes emerge in some `Doomsday scenarios` \cite{Dar1999}.

Strangelet cores themselves have much smaller $Z/A$ than normal nuclei. Core charge is neutralized by leptonic cloud surrounding it, and when the baryon number begin to grow, then after the threshold $A_0\sim10^{15}$  the  core size exceeds the cloud size and the core engulfs the leptons. Regardless the baryon number strangelets are  neutral to outside observers. 
Strangelets could interact with the normal matter in two different ways: if the core charge is positive then the repulsive forces acting on normal nuclei allow only  collisions. This barrier, however, is absent if the core charge is negative -- in  this case colliding normal nuclei would be converted to  strange matter with simultaneous release of large amount of energy. There is no potential barrier for neutron-strangelet interaction, that is why neutron stars could serve as perfect detectors for presence of strangelets in the Galaxy.

Strangelets impinging on the Earth could manifest themselves in different ways \cite{deRujula1984}: energy that is deposited during their passage through medium could be detected in a  host of experiments \cite{BAIKAL,BAKSAN,MACRO,SLIM,ANTARES}. Alternatively, huge exposure of ancient ($\sim10^9$ years) mica crystal \cite{Price1986} or meteorites \cite{Poluhina2013} could be used. These methods  allows to constrain flux of the strangelets in a broad range of their masses.

In this paper  it is suggested to constrain  the strangelet flux using observations of the fluorescence detectors of large cosmic-ray experiments such as the Pierre Auger observatory (PAO) \cite{Prado2005} and the Telescope Array(TA) \cite{Tokuno2012}. Massive strangelets traversing the atmosphere would emit prodigious amount of light \cite{deRujula1984} which in turn could be detected using fluorescence  detectors \cite{Bertaina2014}. Large exposure of these instruments could allow one to obtain competitive constraints on the flux of strangelets.

{\bf 2. Prospects for observations.}

Strangelets are small objects with a density slightly exceeding the nuclear one \cite{Chin1979}:
\be
\rho_s=3.6\times10^{14} \mathrm{g~cm^{-3}}.
\label{eq:dens}
\ee
As stated above, their intrinsic charge is neitralized by leptons. For small strangelets $A<10^{15}$ ($m<1.5$ ng)  this cloud has a fixed radius $r_0\sim10^{-8}$ cm that is  close to  the size of  usual atoms. For more massive strangelets  their neutralizing leptons are hidden inside their cores and their radii begin to grow with the mass:
\be
r(M)=\left(\frac{3M}{4\pi\rho_s}\right)^{1/3}.
\label{eq:radius}
\ee

As a strangelet passes through the Earth's atmosphere it losses its energy through the collisions with the atmosphere nuclei
\be
 dE/dx=-\rho\sigma v^2,
 \label{eq:energylossrate}
 \ee
where $\rho$ is the air density, $\sigma=\pi r^2$ is the cross section and $v$ is the velocity of  the strangelet.
As it would be shown below, only the massive strangelets ($A>10^{20}$) are objects of the primary interest for the proposed approach. Such objects lose negligible fraction of their kinetic energy when travelling through the atmosphere thus $v$ could be regarded as a constant and the total energy loss $\Delta E$ could be straightforwardly calculated:

\be
 \Delta E=\sigma X_{atm} v^2,
 \label{eq:energyloss}
 \ee
where $X=\int \rho dx$ is the column density of the atmosphere, $X_{atm}\sim1000~\mathrm{g~cm^{-2}}$.

These estimates of the energy loss are the lower limits  -- if there are some additional contributions to it than the loss rate would be higher and the constraints could be pushed to smaller cross-sections$/$masses. This technique is also applicable to many other exotic candidates such as Q-balls\cite{Kusenko1998}  or antiquark nuggets \cite{Lawson2011,Gorham2012}, where   additional contributions to the loss rate  could come from catalysis of baryon decay or annihilation. The upper limits on  the energy loss could be estimated as $\sim \sigma X c^2$ and could be larger than ones given by Eq.\ref{eq:energyloss} by 6 orders of magnitude (assuming for the strangelets typical galactic velocities $v\sim 300~ \mathrm{km ~s^{-1}}$). In sharp difference with the case where the energy loss proceeds through collisions only,  a strong emission of high energy particles is expected and that makes possible to detect these objects not only with the fluorescence but with the surface detectors of ultra-high energy cosmic rays (UHECR) observatories as well \cite{Lawson2011}.  
%
The concrete estimate of the projected sensitivity of the technique was done with respect to the Telescope Array. TA is the largest in the Northern hemisphere   UHECR detector located in Utah, USA which has been fully operational since March 2008. It consists of 507 scintillator detectors covering the area of
approximately 700 km$^2$ \cite{Abu-Zayyad2012}, 38 fluorescence telescopes arranged in 3 stations overview the atmosphere over the surface array \cite{Tokuno2012}
The basic idea of the fluorescence observations is rather simple: particles in the electromagnetic component of a  shower caused by a UHECR excite nitrogen molecules. These molecules emit UV-light in the wavelength range 300-400 nm when de-excited,  and this light could be detected on the ground, thus allowing to observe the shower profile.
Though only a minor part ($10^{-4}-10^{-3}$) of the initial energy of the UHECR goes into the fluorescence, this method allows to detect  particles with energies $>10^{18}$ eV at distances larger than 30 km. 

The crucial part of the problem  is the threshold mass of the strangelets that could be detected with this technique. It could be roughly estimated as the mass of the strangelet that produces the same amount of radiation as an $E_0=10^{18}$ eV  UHECR. Energy dissipation in strangelet's case is very close to meteor's, though such processes as ablation are understandably absent. It was shown that luminous effectiveness of meteors lies in $10^{-2}-10^{-1}$ range, depending on velocity and composition of meteors \cite{Ceplecha1998}. Note that it is the \emph{bolometric} effectiveness, and because fluorescence detectors observe in a quite narrow band 300-400 nm, it should be scaled down. As this band resides close to the maximum of the Planck's distribution for temperatures of $\sim$ several eV which are typical for these phenomenae it is robust to use $10^{-2}$ as the  rescaling factor. So it could be adopted that both UHECRs and strangelets have the same luminosity effectiveness, $10^{-4}$ in the relevant passband.

Then, the threshold radius $r_0$ and  mass $M_0$ would be defined as follows

\be
 r_0=\left(\frac{E_0}{\pi X_{atm} v^2}\right)^{1/2}=8\times10^{-7}\left(\frac{300~\mathrm{km~s^-1}}{v}\right)~\mathrm{cm}
 \label{eq:rad_threshold}
 \ee

\bea
 M_0=\frac{4\pi\rho_sr_0^3}{3}= \frac{4\rho_s}{3\sqrt{\pi}}\left(\frac{E_0}{\pi X_{atm} v^2}\right)^{3/2}=\\=6\times10^{-4}\left(\frac{300~\mathrm{km~s^-1}}{v}\right)^3~\mathrm{g}\nonumber
 \eea

The threshold was set at $m_0=6\times10^{-4}~\mathrm{g}\sim4\times10^{20}~\mathrm{GeV}/c^2$ and all derived limits are applicable to strangelets with larger masses. It is worth noting that  at masses $M>9\pi X_{Earth}^3/16\rho_s^2\sim10$ g ($A\sim10^{25}$), where $X_{Earth}= 10^{10}~\mathrm{g~cm^{-2}}$ is the column density of the Earth, strangelets begin to travel through the Earth almost unimpeded and the corresponding field of view of the detectors  approaches $4\pi$.  Also, for more massive strangelets corresponding effective area grows significantly, by factor 5 for $A>10^{23}$ (equivalent to a cosmic ray with the energy $\sim10^{19.5}$ eV)

The TA fluorescence detectors are adapted for the observations of the UHECRs which travel at the speed of light: they are triggered if there is a signal in  $\ge5$ adjacent photo-multiplier tubes (PMT) in 25.6 $\mu s$ time window, each PMT having a field of view $\sim1^{\circ}$\cite{Tameda2009}. Strangelets are supposed to travel with non-relativistic velocities, $v\sim10^{-3}c$, so they  cannot trigger the instrument by themselves. 
However, in addition to instant radiation from the strangelet one would expect bright meteor-like wake radiation,  that  lasts $10^{-3}-10^{-1}$ s \cite{Ceplecha1998}.  This wake emission would be observed as short trail after the strangelet and would trigger the detectors.

The limits on the flux of the strangelets that could be potentially detected with this technique could be readily estimated. 

\be
\Phi\sim\frac{1}{2\pi\mathcal{A}t}=5\times10^{-22}~\mathrm{cm^{-2} s^{-1} sr^{-1}},
 \label{eq:flux}
 \ee
 where $\mathcal{A}\sim10^{3}~\mathrm{km^2}$ is the effective area of TA FD at energies $10^{18}$ eV \cite{Tameda2009}, and  $t=3\times10^{7}$ s is the total duration of the observations taking into account their $10-12\%$ duty cycle.
 
It should be mentioned that the data from  the fluorescence detectors of the PAO \cite{Prado2005} could be used   in this technique as well and that could lead to even  deeper limits, taking into account longer duration of observations and  the  effective erea of the FD detectors that is larger by factor 2.

Finally, though there is a  possibility that there exist meteors moving with 'galactic' velocities \cite{Afanasiev2007}, strangelets could be fairly easily distinguished from them using difference in their trail profiles: usual meteors disintegrate at altitudes higher than around 100 km, when the  strangelets would travel all the way down to the surface. 

{\bf 3. Conclusions.}
The proposed technique could be used to constrain flux of strangelets (also, Q-balls of the strong interacting variety, antiquark nuggets and related exotica) impinging upon the Earth. For strangelets with baryon number $A>4\times10^{20}$ it is possible to reach limit of $\Phi\sim5\times10^{-22}~\mathrm{cm^{-2} s^{-1} sr^{-1}}$, the limits are stronger by an order of magnitude  for $A>10^{25}$.  These limits would improve by 1.5 (2.5) orders of magnitude the strongest present ones obtained from ancient mica crystals.

\paragraph*{Acknowledgements.} 
The author wants to thank S. Troitsky, G. Rubtsov, and O. Kalashev for valuable comments.
The work of the author is supported by  the Russian Science Foundation grant 14-12-01340.  The author acknowledges the fellowship of the Dynasty Foundation
\bibliography{strangelet}

%


\end{document}